%% file: paper.tex
\documentclass{juliacon}
\usepackage{amsmath}
\usepackage{booktabs}

\graphicspath{{assets/}}

\begin{document}

\input{header}

\maketitle

\begin{abstract}

Tensor Networks are graph representations of summation expressions in which vertices represent tensors and edges represent tensor indices or vector spaces. In this work, we present \texttt{EinExprs.jl}, a Julia package for contraction path optimization that offers state-of-art optimizers. We propose a representation of the contraction path of a Tensor Network based on symbolic expressions. Using this package the user may choose among a collection of different methods such as Greedy algorithms, or an approach based on the hypergraph partitioning problem. We benchmark this library with examples obtained from the simulation of Random Quantum Circuits (RQC), a well known example where Tensor Networks provide state-of-the-art methods.

\end{abstract}

\section{Introduction}
A Tensor Network is a collection of tensors connected by common indices indicating contraction operations. Despite being extensively used in different fields of Physics such as Quantum Information \cite{evenbly2022practical} or Condensed Matter\cite{SCHOLLWOCK201196}, recently they have received much attention due to their capabilities to simulate Quantum circuits, a task hard even for supercomputers \cite{PhysRevX.10.041038}. The necessity to improve Tensor Network methods emerges from the computational resources required to manipulate these structures. Currently, Tensor Network methods are state of the art for quantum circuit simulation.

Tensor Networks are equivalent to a graph representation of Einstein summation expressions (a.k.a. \textit{einsum}) in which vertices represent tensors and edges represent tensor indices or vector spaces. A tensor $T$ is encoded by the Tensor Network, and can be exactly computed by contracting the tensors following the summation operations. As an example, using Einstein summation rules for common indices, $T$ is the result of the contraction of several Tensors:

\begin{equation}\label{eq:tn-example}
    T = A_{im} B_{ijp} C_{jkn} D_{klp} E_{mno} F_{lo} 
\end{equation} 

\begin{figure}[h]
    \centering
    \includegraphics[width=\columnwidth]{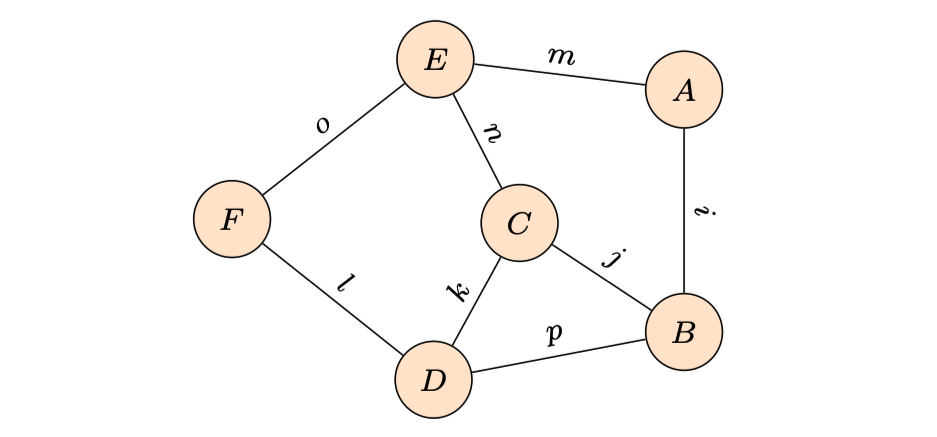}
    \caption{Diagrammatic representation of Equation~\ref{eq:tn-example} as a Tensor Network.}
    \label{fig:tn-example}
\end{figure}

Any \textit{einsum} expression can be reinterpreted diagrammatically using Tensor Networks. For example, Equation~\ref{eq:tn-example} is represented graphically as shown in Figure~\ref{fig:tn-example}. The order in which the tensors are contracted highly affects the computational cost of the simulation. Indeed, exact tensor network contraction is a \#P-complete problem~\cite{garcia2012exact}. Finding the optimal contraction path of a tensor network is known to be linked to the optimal tree decomposition problem of the underlying graph \cite{markov2008simulating}. This is equivalent to finding the treewidth of such graph, which is a well-known NP-complete problem.

In this work we present \texttt{EinExprs.jl}, a package forTensor Network contraction path optimization and visualization. Many of the Tensor Networks found in the literature have some kind of structure. As with many NP-complete problems, this structure can be exploited to reduce the complexity of the problem. Recent developments in the field have demonstrated that some heuristics are well-suited for finding quasi-optimal contraction paths. \texttt{EinExprs.jl} aims to be the reference package for the development of new algorithms by providing an easy interface along the fastest implementations of well-known algorithms.

The work is organized as follows. Section~\ref{sec:related-work} reviews state-of-art software for contraction path optimization. Section~\ref{sec:symbolic-expr} relates contraction paths and symbolic expressions. Section~\ref{sec:optimization-methods} introduces some of the most popular contraction path optimization methods which are implemented in \texttt{EinExprs}. Section~\ref{sec:benchmarks} compares the execution performance of \texttt{EinExprs} against other packages.

\section{Related work}\label{sec:related-work}
The reference software package to provide a diverse set of contraction path optimizers is \texttt{opt\_einsum} \cite{daniel2018opt}. It provides implementations of depth-first exhaustive search, greedy search and some tree-width estimation algorithms from the graph theory world such as QuickBB~\cite{gogate2012complete} or dynamic programming. A new optimizer based on the well-known Hypergraph Partitioning problem was presented in \cite{gray2021hyper}. Along with hyper-parameter optimization, this work pushes state-of-art contraction path optimization for large-scale tensor networks. 

Many packages provide support for einsum-like notation in Julia: \texttt{TensorOperations.jl}~\cite{devos2023tensoroperations}, \texttt{ITensors.jl}~\cite{fishman2022itensor}, \texttt{OMEinsum.jl}~\cite{omeinsum} and \texttt{Tullio.jl}~\cite{tullio}, to name a few.
The \texttt{TensorOperations} \cite{devos2023tensoroperations} package provides the fastest implementation of the exhaustive optimizer \cite{pfeifer2014faster}.
The recent \texttt{OMEinsumContractionOrders.jl} reimplements in Julia some of the algorithms found in \texttt{opt\_einsum} together with some of their own, and currently powers \texttt{OMEinsum.jl} and \texttt{ITensorNetworks.jl}.

\section{Contraction paths are symbolic expressions}\label{sec:symbolic-expr}

Working with large Tensor Networks involves choosing the best data-structures to avoid unwanted overheads when scaling up. The same can be said for contraction paths. \texttt{opt\_einsum} stores contraction paths as an ordered list of pairs of SSA ids of tensors, which is equivalent to storing the contracting indices of the represented pairwise tensor contraction. We argue that such data-structure does not fully exploit the structure inherent in contraction paths, and that a better representation can be attained.

It is important to observe that indices in a contraction path follow a partial order, represented as $(\alpha_1 \ldots \alpha_k)$ indicating the precedence in the contraction of indices $\alpha_1 \ldots \alpha_k$. As an example, in Figure~\ref{eq:tn-example} it can be easily checked that $(m,o,j,pk,inl)$, $(m,j,o,pk,inl)$, $(m,j,pk,o,inl)$, $(j,pk,m,o,inl)$, $(j,m,pk,o,inl)$ and $(j,m,o,pk,inl)$ generate exactly the same intermediate tensors, where summation indices are explicitly indicated:
\begin{align*}
\label{eq:contraction-path}
    \alpha_{ino} &= \sum_{m} A_{im} E_{mno} \\
    \beta_{inl} &= \sum_{o} \alpha_{ino} F_{ol}\\
    \gamma_{ipkn} &= \sum_{j} B_{ijp} C_{jkn} \\
    \delta_{inl} &= \sum_{pk} \gamma_{ipkn} D_{pkl}\\
    T &= \sum_{inl} \beta_{inl} \delta_{inl} 
\end{align*} 

In search for a better suited data-structure and inspired by Julia's LISP heritage, we draw a parallelism between symbolic expressions and contraction paths.
In a symbolic expression, a code expression is decomposed in a syntax tree: The terminal nodes or \textit{leaves} represent the initial values or variables which the computation takes as inputs, while non-terminal nodes or \textit{branches} represent computations. The tree diagram of a syntax tree faithfully represents the partial order implicit in the symbolic expression: the sequential order of execution is unfixed and free to reconfigure as long as the precedence set by the tree is respected. In order to allow composition of expressions, symbolic expressions are usually implemented as recursive data-structures: a symbolic expression stores a symbol (i.e. the name of the performed operation) and a list of other symbolic expressions.

\begin{figure}
    \centering
    \includegraphics[width=\columnwidth]{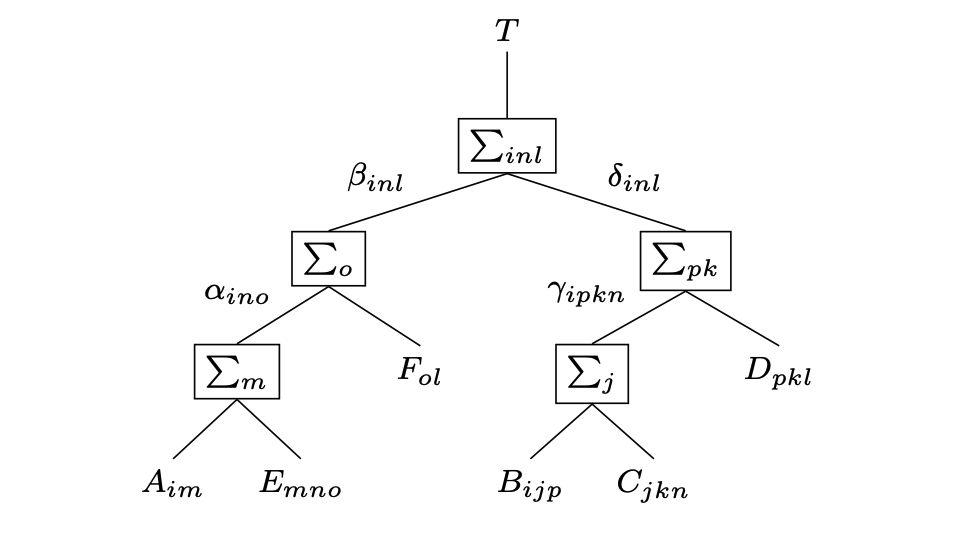}
    \caption{A valid contraction tree of the Tensor Network found in Figure~\ref{fig:tn-example}. Vertices represent contraction operations, open edges represent initial tensors (except for the final tensor $T$) and closed edges represent intermediate tensors.}
    \label{fig:contraction-tree}
\end{figure}

We use a similar approach for Tensor Network contractions. Following the example from Figure~\ref{eq:tn-example}, we observe that the contraction path $(m,o,j,pk,inl)$, and the equivalent contraction paths, can be represented as a tree diagram in Figure~\ref{fig:contraction-tree}, where vertices represent contraction operations, open edges represent initial tensors (except for the final tensor $T$) and closed edges represent intermediate tensors. Note that with this representation, the particular order on which partial contraction operations are performed is not explicitly specified.

\begin{figure}
    \centering
    \includegraphics[width=\columnwidth]{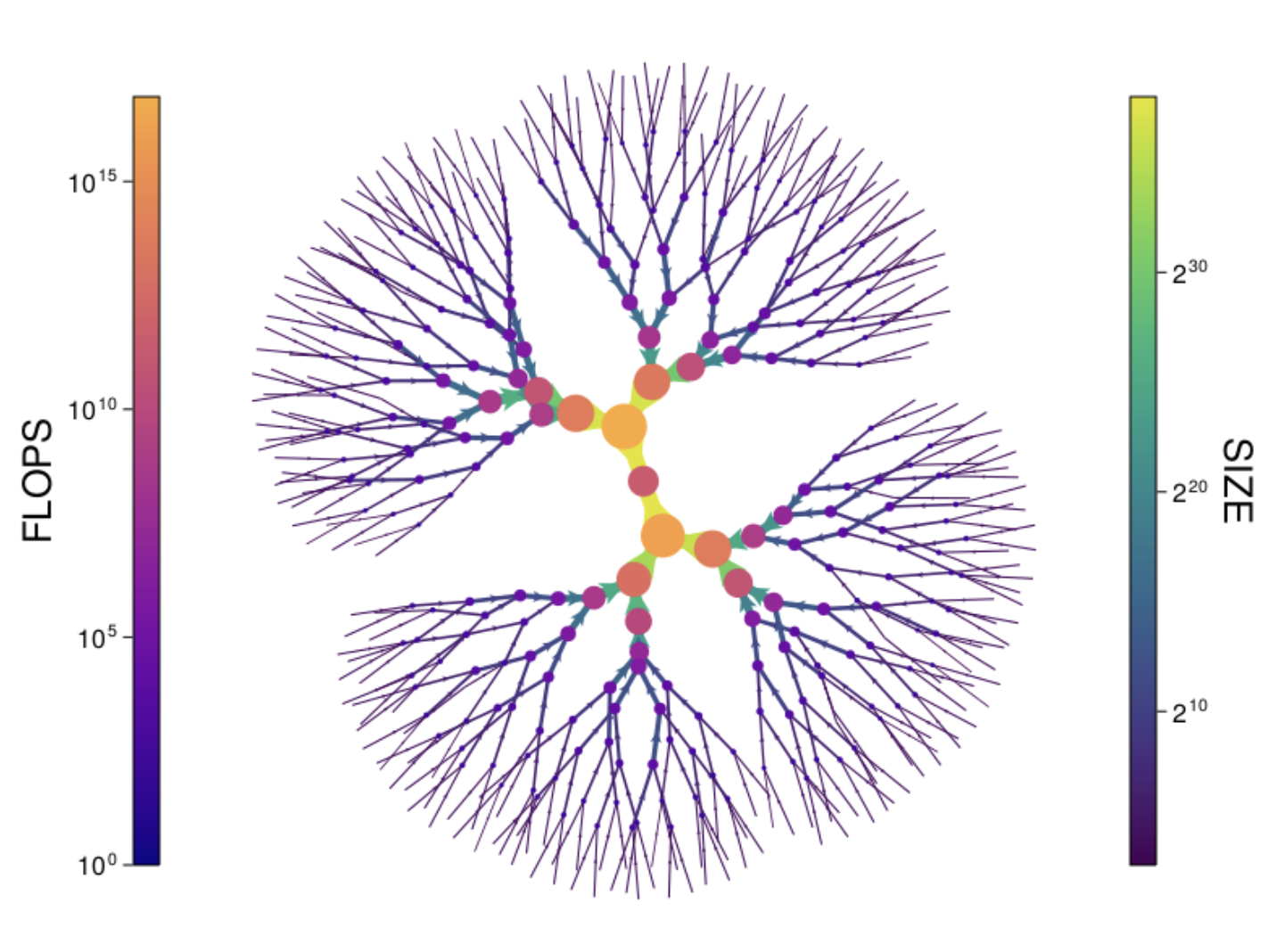}
    \caption{Illustration of a contraction tree using \texttt{Makie.jl}. Vertices represent tensor contraction operations, with larger sizes and brighter colors for contractions with larger amount of FLOPs. Edges represent intermediate tensors, with larger thickness and brighter colors for larger amount of memory used.}
    \label{fig:contraction-tree:makie}
\end{figure}

In such tree visualization, computational cost information can be mapped to the nodes and edges. 
\texttt{EinExprs} integrates with the \texttt{Makie}~\cite{DanischKrumbiegel2021} library to add plotting capabilites of the contraction trees.
In Figure~\ref{fig:contraction-tree:makie}, we plot the contraction tree of a Random Quantum Circuit.
The size and color of the tree nodes are related to the cost in FLOPs of the tensor contractions, and the thickness and color of the edges are related to size of the intermediate tensors.
It can be easily seen that most tensor contraction operations are of negligible size and that only a few tensor contractions are of a large size.
This suggests that in this case, the cost of contracting the Tensor Network is dominated by the intermediate tensors in the last steps.

\section{Optimization methods for contraction paths}\label{sec:optimization-methods}

We can distinguish two kinds of optimizers based on the starting point of the construction of the contraction tree: local optimizers, which start from the leaves up to the root; and global optimizers, which start from the root down to the leaves. In this section, we present the collection of methods implemented in \texttt{EinExprs.jl}, with a description of their advantages and limitations.

\subsection{Exhaustive}
The Exhaustive search explores the full combinatorially-big solution space.
As such, it guarantees to find the optimal contraction path but at a factorial complexity $\mathcal{O}(n!)$. The complexity can be relaxed down to $\mathcal{O}(e^n)$ by excluding outer products, which rarely obtain any gain.

\texttt{EinExprs.jl} provides two implementations: a depth-first exhaustive search based on the \textit{Optimal} optimizer found in \texttt{opt\_einsum}, which uses backtracking along with path pruning. Also, a breadth-first exhaustive search based on the implementation of \texttt{TensorOperations.jl}~\cite{pfeifer2014faster}.

\subsection{Greedy}
The Greedy algorithm works by \textit{greedily} selecting the contracting index that maximizes a given score, which is not directly linked with the cost of contraction but instead it follows a heuristic. The most common score heuristic evaluations use the size of the resulting tensor subtracted by the size of the input tensors.

The Greedy algorithm is an extremely fast local optimizer, although its results are far from optimal on large networks. Luckily, these results can be improved by adding a thermal noise to the greedy candidate selector and sampling from it many times.

\subsection{Hypergraph Partitioning}

Treewidth calculation is a persistent problem in the community detection field.
Over the years, researchers have developed many algorithms and heuristics for indirectly finding the treewidth of particular graph instances.
It is then presumable that community detection methods could perform well on contraction path optimization.
Based on this premise, authors in \cite{gray2021hyper} formulated the contraction path optimization problem as a Hypergraph Partitioning problem, a well-known and largely worked out community detection problem.
The problem consists of dividing the vertices of a (hyper)graph into 2 sets such that the number of edges in common between the sets is minimized.
The parallelism with a global contraction path optimizer is obvious: the 2 vertex sets would be the topmost intermediate tensors in the contraction tree and the shared edges between the 2 sets are the contracting indices.
This method is then called recursively. Moreover, as a global optimizer it can easily be composed with other local optimizers.
For example, when achieving a small enough vertex set, the problem could be forwarded to a Exhaustive optimizer for optimal solution of that subproblem. 


\section{Benchmarks}\label{sec:benchmarks}
We analyze the performance of \texttt{EinExprs} using as a reference the package \texttt{TensorOperations} for the Exhaustive search, and \texttt{OMEinsumContractionOrders} for the Greedy algorithm.
Random Tensor Networks are used in both experiments.
For the Exhaustive search experiment, we set some constraints such as the maximum number of contracting indices (32), the maximum number of open indices (10) and the maximum size of an index (5).
For the Greedy algorithm experiment, we fix the number of initial tensors to powers of 2, until a maximum of 4096, and the "regularity" (or average number of contracting indices per tensor) to 3.
In our analysis, error bars are not shown due to small variance of the results.
Benchmarks for Hypergraph Partitioning were skipped due to all libraries being powered by \texttt{KaHyPar}~\cite{10.1145/3529090}.
Execution time benchmarks were carried out in a single-core of the Marenostrum 4 supercomputer at the Barcelona Supercomputing Center.

\begin{figure}
    \centering
    \includegraphics[width=\columnwidth]{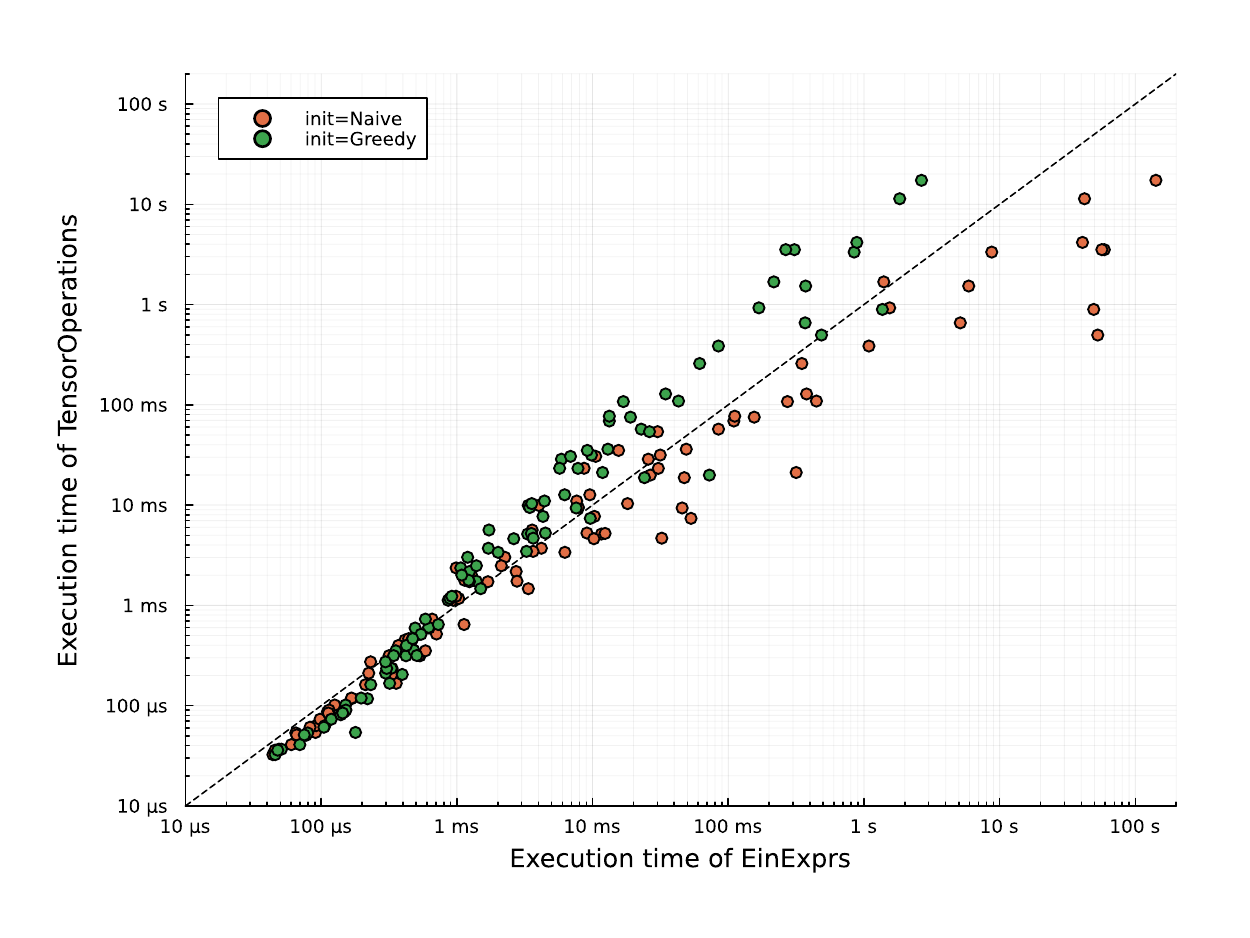}
    \caption{Comparison of execution time between \texttt{EinExprs} ($X$-axis) and \texttt{TensorOperations} ($Y$-axis) implementations of the breadth-first exhaustive optimizer. Samples above the dashed line where solved faster with \texttt{EinExprs} than with \texttt{TensorOperations}, and vice-versa. Orange samples had no initial estimation of the contraction cost, while green samples were initialized with the Greedy optimizer.}
    \label{fig:exhaustive:time}
\end{figure}

In Figure~\ref{fig:exhaustive:time}, we compare the execution time of the breadth-first exhaustive optimizer implementations of \texttt{EinExprs} against \texttt{TensorOperations}, over random tensor networks of up to 32 tensors.
The dashed line represents equal time for both implementations. One of the advantages of \texttt{EinExprs} over \texttt{TensorOperations} is its capability to compose different optimizers together. By performing a initial estimation of the computational cost with the greedy optimizer, the path pruning can act before and filter out more paths. It can be seen that this initial estimation (labeled \textit{init=Greedy}) speeds up the optimization by up to 2 orders of magnitude compared to no initial estimation (labeled \textit{init=Naive}).
For small tensor networks, both implementations take a similar amount of time.
For larger tensor networks, the \texttt{EinExprs} implementation without estimation lacks behind \texttt{TensorOperations} but the initial estimation approach compensates the loss of performance and accelerates over it.
This suggests that even though our implementation of the breadth-first Exhaustive search is not as optimized as \texttt{TensorOperation}'s implementation, the initial estimation approach is algorithmically superior.

\begin{figure}
    \centering
    \includegraphics[width=\columnwidth]{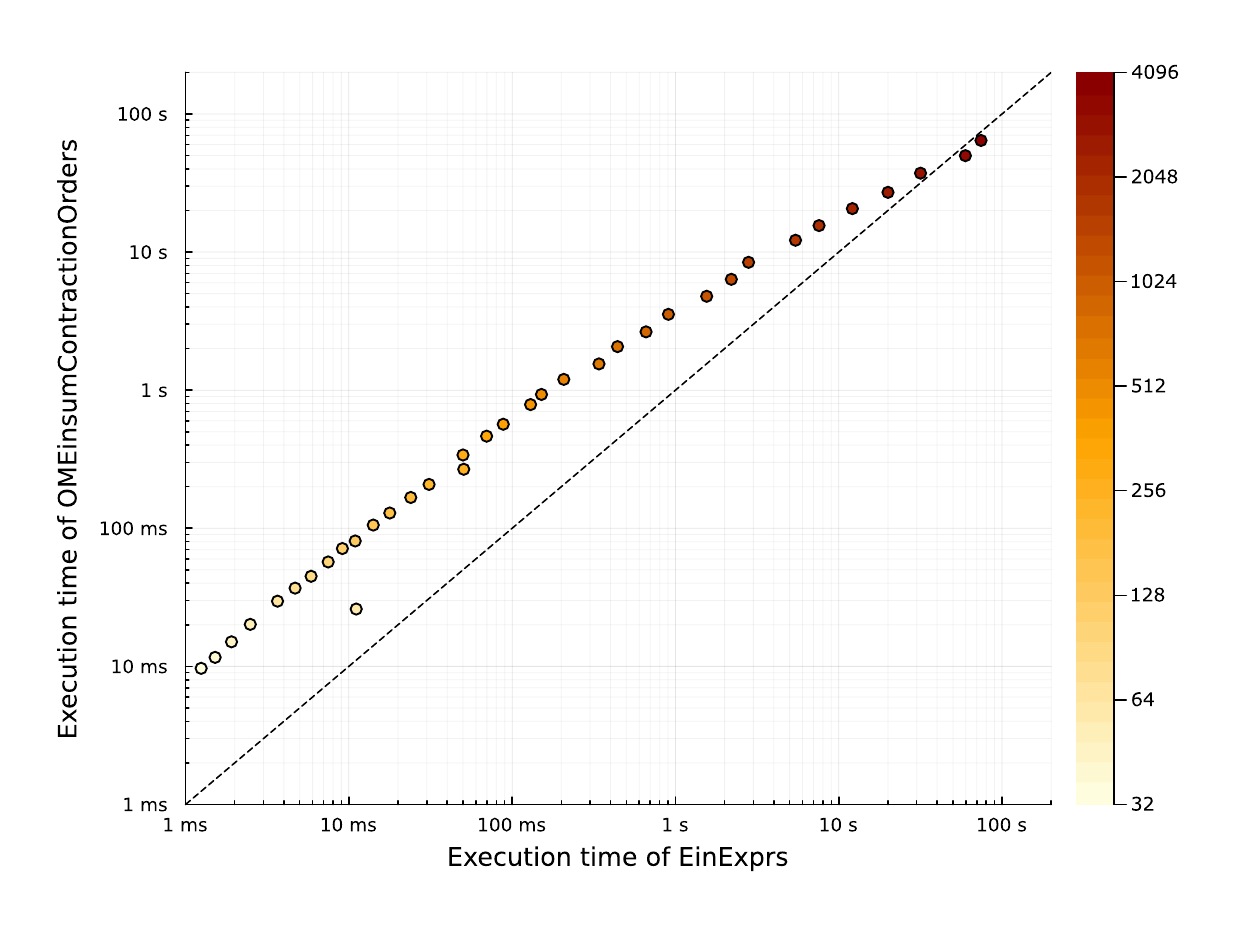}
    \caption{Comparison of execution time between \texttt{EinExprs} ($X$-axis) and \texttt{OMEinsumContractionOrders} ($Y$-axis) implementations of greedy optimizer. Samples above the dashed line where solved faster with \texttt{EinExprs} than with \texttt{OMEinsumContractionOrders}, and vice-versa. The color indicates the number of tensors.}
    \label{fig:greedy:time}
\end{figure}

In Figure~\ref{fig:greedy:time}, we compare the execution time of the greedy optimizer implementation of \texttt{EinExprs} against the implementation of \texttt{OMEinsumContractionOrders}.
\texttt{EinExprs} consistently achieves 1 order of magnitude speedup on tensor networks of up to 512 tensors.
On larger tensor networks, the time difference progressively vanishes until achieving a similar runtime on tensor networks of around 4000 tensors.
Although the reason behind this behavior is not fully clear yet, we hypothesize that it is due to an algorithmic overhead.
In particular, we use a heap for storing candidate contractions that needs to be updated on each winner selection.
On worst-case scenario, updating a candidate in the heap involves $\mathcal{O}(\log n)$ operations and as the size of the problem grows, the depth of the heap and the probability of needing a larger amount of operations on the update grows along.
An alternative hypothesis that we also consider is L1-cache saturation.
An argument that supports this idea is that the speedup between both implementations is consistent until a size of around 512 tensors, where the heap no longer fits in the L1 cache (32~KiB).
In any case, more work is needed to elucidate the reason behind the performance loss.

\section{Summary}

\texttt{EinExprs} is a Julia package for Tensor Network contraction path optimization. It defines a data structure for the description of the contraction path, and allows the user to achieve top performance by the combination of the optimization methods implemented. It achieves up to 1 order of magnitude speedups compared to other popular packages, and It also offers a carefully crafted design for user experimentation and code composition. It currently powers some of the large-scale quantum tensor network simulations performed at Barcelona Supercomputing Center, with applications to Quantum Circuit simulations.

In the future, perspective work on the package includes addressing the following features:
\begin{itemize}
    \item Implement missing path optimizers from \texttt{opt\_einsum}, such as QuickBB or Dynamic Programming.
    \item Optimization of tensors with asymptotic scaling.
    \item Implement post-optimizers such as subforest reconfiguration optimization or index permutation for minimization of memory accesses.
\end{itemize}

The source code of the package can be found in \url{https://github.com/bsc-quantic/EinExprs.jl}.

\section{Acknowledgements}

Authors would like to thank Lukas Devos and Jutho Haegeman for discussions on performance and testing. Also, Joseph Tindall and Matthew Fishman for their help on integrating \texttt{EinExprs} in \texttt{ITensorNetworks}.

\bibliographystyle{juliacon}
\bibliography{ref.bib}

\end{document}

%% file: header.tex

\title{\texttt{EinExprs}: Contraction Paths as Symbolic Expressions}

\author[1]{Sergio Sanchez-Ramirez}
\author[1]{Jofre Vallès-Muns}
\author[1, 2]{Artur Garcia-Saez}
\affil[1]{Barcelona Supercomputing Center, 08034 Barcelona, Spain}
\affil[2]{Qilimanjaro Quantum Tech., 08014 Barcelona, Spain}

\keywords{Julia, Tensor Networks, Contraction Path, Symbolic Expressions, Optimization}